\documentclass[aps,pra,reprint]{revtex4-1}

\usepackage{amsmath}    
\usepackage{graphicx}   
\usepackage{hyperref}   
\usepackage{textcomp}   
\usepackage{bm}         
\usepackage{siunitx}    

\newcommand*{\bvec}[1]{\bm{#1}}
\newcommand*{\slevel}{\smash{\ensuremath{\text{6s}\,^\text{2}\text{S}_\text{1/2}}}}
\newcommand*{\plevel}{\smash{\ensuremath{\text{6p}\,^\text{2}\text{P}_\text{1/2}}}}
\newcommand*{\dlevel}{\smash{\ensuremath{\text{5d}\,^\text{2}\text{D}_\text{3/2}}}}
\newcommand*{\sterm}{\smash{\ensuremath{^\text{2}\text{S}_\text{1/2}}}}
\newcommand*{\pterm}{\smash{\ensuremath{^\text{2}\text{P}_\text{1/2}}}}
\newcommand*{\dterm}{\smash{\ensuremath{^\text{2}\text{D}_\text{3/2}}}}
\newcommand*{\nudp}{\smash{\ensuremath{\nu_{(\text{5d}\,^\text{2}\text{D}_\text{3/2}\,\text{--}\,\text{6p}\,^\text{2}\text{P}_\text{1/2})}}}}
\newcommand*{\nusp}{\smash{\ensuremath{\nu_{(\text{6s}\,^\text{2}\text{S}_\text{1/2}\,\text{--}\,\text{6p}\,^\text{2}\text{P}_\text{1/2})}}}}
\newcommand*{\nusd}{\smash{\ensuremath{\nu_{(\text{6s}\,^\text{2}\text{S}_\text{1/2}\,\text{--}\,\text{5d}\,^\text{2}\text{D}_\text{3/2})}}}}

\DeclareSIUnit\count{cnt}

\begin{document}

\title{Determination of transition frequencies in a single \texorpdfstring{$^\text{138}\text{Ba}^\text{+}$}{138Ba+} ion}
\date{\today}
\author{E. A. Dijck}
\email{e.a.dijck@rug.nl}
\author{M. \surname{Nu\~nez Portela}}
\author{A. T. Grier}
\altaffiliation[Present address: ]{Department of Physics, Columbia University, New York, NY 10027}
\author{K. Jungmann}
\author{A. Mohanty}
\author{N. Valappol}
\author{L. Willmann}
\affiliation{Van Swinderen Institute, University of Groningen, The~Netherlands}

\begin{abstract}
Transition frequencies between low-lying energy levels in a single trapped $^\text{138}\text{Ba}^\text{+}$ ion
have been measured with laser spectroscopy referenced to an optical frequency comb.
By extracting the frequencies of one-photon and two-photon components of the line shape using an eight-level optical Bloch model,
we achieved \SI{0.1}{\mega\hertz} accuracy for the \dlevel{} -- \plevel{} and \slevel{} -- \dlevel{} transition frequencies,
and \SI{0.2}{\mega\hertz} for the \slevel{} -- \plevel{} transition frequency.
\end{abstract}

\maketitle


Trapped single ions can be exploited to investigate the interaction 
between light and matter, and to construct optical clocks~\cite{*[{}][{ and references therein.}]Ludlow2015}.
For these applications based on high precision spectroscopy
a good understanding of the optical line shapes involved is indispensable.
We have employed an optical frequency comb~\cite{Hall2006,*Haensch2006} to measure
transition frequencies in $\text{Ba}^\text{+}$ and a model based on optical Bloch equations to extract atomic parameters from fluorescence spectra.
These are essential ingredients for high precision experiments,
in particular for atomic parity violation measurements in single $\text{Ba}^\text{+}$~\cite{Fortson1993,Koerber2003} and 
$\text{Ra}^\text{+}$ ions~\cite{Wansbeek2008,NunezPortela2014} in the search for new physics~\cite{Kumar2013}.

\begin{figure}
\begin{center}
  \includegraphics{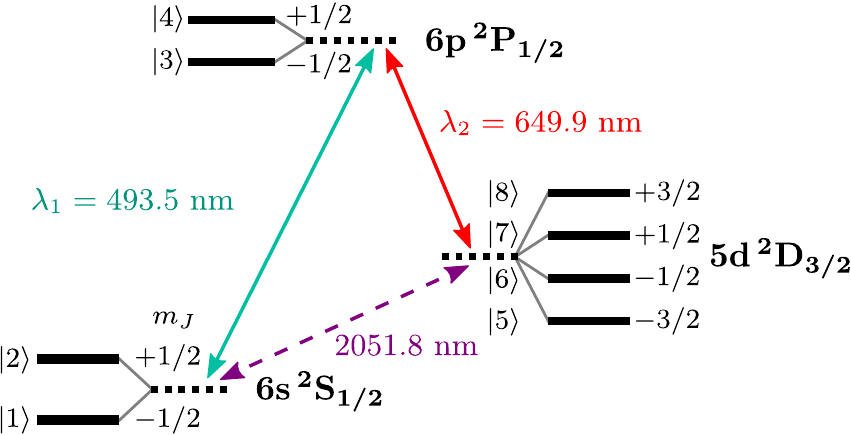}
  \caption{Low-lying energy levels of the $\text{Ba}^\text{+}$ ion. The 
wavelengths of the investigated transitions are given.}
  \label{Fig:BaLevels}
\end{center}
\end{figure}

In this work the frequencies of transitions between three of the lowest fine structure 
levels in the $^\text{138}\text{Ba}^\text{+}$ ion are addressed. 
These levels form a $\Lambda$-configuration as shown in Fig.~\ref{Fig:BaLevels}.
The level \plevel{} decays to the levels \slevel{} and \dlevel{} with a branching ratio of about 
3:1~\cite{DeMunshi2014}. We have measured the transition frequencies in a single $^\text{138}\text{Ba}^\text{+}$ 
ion by driving the transitions \slevel{} -- \plevel{} and \dlevel{} -- \plevel{}, employing an optical frequency comb as frequency reference.
The dynamics of the population of the \pterm{} level can be described by optical Bloch equations~\cite{Stalgies1998,Oberst1999,Zanon-Willette2011}.
Coherent coupling between the \sterm{} and \dterm{} levels is observed when the two laser 
fields are detuned by the same amount from the respective atomic resonances.
In this condition a two-photon process causes coherent population trapping, reducing the population 
of the \pterm{} level~\cite{Siemers1992}.

\begin{figure}
\begin{center}
  \includegraphics{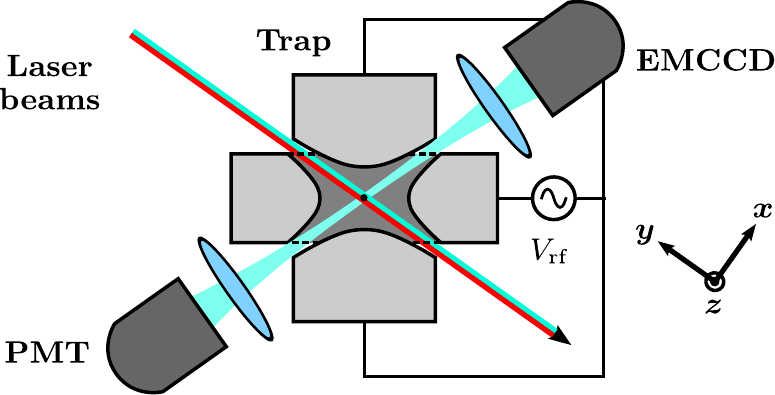}
  \caption{Schematic diagram of the hyperbolic Paul trap used for trapping 
$\text{Ba}^\text{+}$ ions, consisting of a ring electrode and two end caps.
The inner diameter of the ring is \SI{5}{\milli\meter}. Light scattered by ions is detected with a photomultiplier 
tube (PMT) and an electron-multiplying CCD camera (EMCCD).}
  \label{Fig:Trap}
\end{center}
\end{figure}

For the measurements reported here a single $\text{Ba}^\text{+}$ ion is confined in a hyperbolic 
Paul trap (see Fig.~\ref{Fig:Trap}). The trap 
is operated at frequency $\omega_\text{rf} / 2\pi = \SI{5.44}{\mega\hertz}$ with a peak-to-peak
rf~voltage of typically $V_\text{rf} = \SI{600}{\volt}$. Additional electrodes provide a dc potential to compensate 
the effect of mechanical imperfections and stray fields, minimizing the micromotion of 
the ion in the trap. The trap is loaded by photoionization of 
$^\text{138}\text{Ba}$ atoms with laser light at wavelength \SI{413.6}{\nano\meter}. 
The trap is mounted in a UHV chamber with residual gas pressure below \SI{e-10}{\milli\bar}.

Doppler cooling and detection of the $\text{Ba}^\text{+}$ ions is achieved with 
laser light at wavelengths $\lambda_1$ and $\lambda_2$ (see Fig~\ref{Fig:BaLevels}). Laser light to 
drive the \slevel{} -- \plevel{} transition (at wavelength~$\lambda_1$) is generated by frequency doubling
light at wavelength \SI{987}{\nano\meter} from a single-frequency Ti:sapphire laser (Coherent~MBR-110)
in a temperature-tuned MgO:PPLN crystal (Covesion MSHG976-0.5) operated at \SI{156.8}{\degreeCelsius} in 
a linear enhancement cavity. Light to drive the \dlevel{} -- \plevel{} 
transition (at wavelength~$\lambda_2$) is generated by a ring dye laser (Coherent CR-699) operated with DCM 
(4-(dicyanomethylene)-2-methyl-6-(4-dimethylaminostyryl)-4H-pyran),
pumped by a solid-state laser at \SI{532}{\nano\meter} (Coherent Verdi~V10).

The laser beams at wavelengths $\lambda_1$ and $\lambda_2$ are delivered to 
the trap via one single mode optical fiber (Thorlabs PM460). The 
polarization is controlled with linear polarizers and half-wave 
plates. The bichromatic light is focused to a beam waist 
of \SI{60}{\micro\meter} at the trap center with a set of achromatic lenses. 
Typical laser light intensities at the position of the ion are of order 
saturation intensity for $\lambda_1$ and of up to four times saturation intensity for $\lambda_2$. A magnetic 
field of \SI{170}{\micro\tesla} breaks the degeneracy of the magnetic sublevels.
The direction and magnitude of the magnetic field are controlled with 
three pairs of coils.

$\text{Ba}^\text{+}$ ions are detected via fluorescence of the 
\plevel{} -- \slevel{} transition. The light is imaged onto a photomultiplier 
tube (Hamamatsu H11123) and onto an electron-multiplying CCD camera (Andor iXon), see Fig~\ref{Fig:Trap}. 
Background light is suppressed by a band-pass filter with \SI{10}{\nano\meter} bandwidth and 
$>85\%$ transmission at wavelength $\lambda_1$ (Edmund Optics \#65-148).
Light for the PMT is collected within a solid angle of \SI{0.03}{\steradian};
the count rate is of order \SI{2e3}{\count/\second} for a single ion with both 
laser fields on resonance. Light is imaged 
onto the EMCCD camera with a $16\times$ magnifying telescope. The camera provides for the 
observation of localized single ions and crystals formed by several ions.
The size of the image of an ion is proportional to the amplitude of its motion 
and gives an upper limit for its temperature. For a single laser-cooled 
ion this limit is of order $\leq \SI{10}{\milli\kelvin}$ ($\approx \SI{1}{\meter/\second}$);
the Doppler limit for the $\text{Ba}^\text{+}$ cooling transition is $\SI{0.5}{\milli\kelvin}$.

\begin{figure}
\begin{center}
  \includegraphics{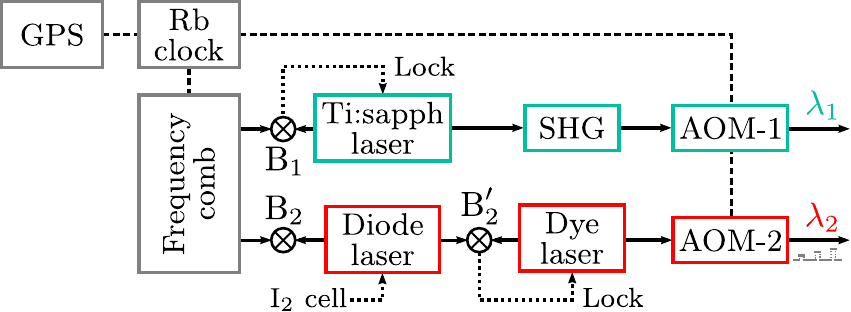}
  \caption{Scheme to transfer the $\Delta\nu/\nu = 10^{-11}$ frequency stability of the 
frequency comb to light at wavelengths $\lambda_1$ and $\lambda_2$.
The light at wavelength $\lambda_1$ is produced by second harmonic generation (SHG)
of light from a Ti:sapphire laser, locked to the frequency comb with beat note $\text{B}_1$.
The light at wavelength $\lambda_2$ is referenced to the frequency comb
via a diode laser stabilized to $\text{I}_\text{2}$ line 
$\text{P}(25)\;6\text{--}5\;a_3$. The frequency of this laser is measured 
with beat note $\text{B}_2$. The light at wavelength $\lambda_2$ itself is produced with a dye laser
offset-locked to the reference diode laser with beat note $\text{B}'_2$.
AOM-2 is switched between two frequencies for laser cooling and for probing the signal.}
  \label{Fig:LaserLock}
\end{center}
\end{figure}

A fiber-based femtosecond frequency comb (Menlo Systems FC1500) is used to measure the frequency
of the light at wavelengths $\lambda_1$ and $\lambda_2$ (see Fig~\ref{Fig:LaserLock}).
Its long term stability is provided by a GPS disciplined rubidium clock (FS 725) 
with an intrinsic frequency stability of $10^{-11}$ in \SI{1}{\second} integration time and 
$10^{-12}$ in \SI{10000}{\second}. This clock also serves as long term reference
for all rf frequencies in the experiment and its stability has been verified to $\Delta\nu/\nu \approx 10^{-11}$ 
via an optical fiber network~\cite{Pinkert2015}.

The frequency of the light at wavelength $\lambda_1$ is determined by counting beat note frequency $\nu_{\text{B}_1}$
between light from a frequency comb mode and light from the Ti:sapphire laser,
which is actively stabilized (see Fig.~\ref{Fig:LaserLock}).
Varying the frequency of the light at wavelength $\lambda_1$ was achieved by changing the repetition rate of the comb.
After frequency doubling, the light at wavelength $\lambda_1$ passes through an acousto-optic modulator operated at fixed frequency $\nu_\text{AOM-1}$.
The frequency of the second light field at wavelength $\lambda_2$ is determined via an intermediate diode laser,
the short term frequency stability of which is provided by saturated absorption spectroscopy~\cite{Dammalapati2009} of 
component $a_3$ of line $\text{P}(25)\;\text{6--5}$ in molecular $\text{I}_\text{2}$.
Its frequency is determined by counting beat note frequency $\nu_{\text{B}_2}$ with light from the frequency comb.
This setup has also provided the frequency of the line as $\nu_{\text{P}(25)\;6\text{--}5\;a_3} = \SI{461312288.10(2)}{\mega\hertz}$
at an $\text{I}_\text{2}$ cell temperature of \SI{25}{\degreeCelsius}, corresponding to a vapor pressure of \SI{0.4}{\milli\bar}.
This particular line was not calibrated to this accuracy before in a compilation~\cite{Gerstenkorn1980,Xu2000}.
The dye laser producing the light at wavelength $\lambda_2$ is locked to the reference diode laser at a variable offset 
using an additional beat note at frequency $\nu_{\text{B}'_2}$.
The light is then frequency-shifted by double passing through an acousto-optic modulator.
This device is switched at \SI{53.6}{\kilo\hertz} between two frequency settings $\nu_\text{AOM-2}$.
One frequency is variable and serves for probing the transition of interest for \SI{3.7}{\micro\second}.
The second setting provides a fixed frequency for laser cooling the ion during the remaining \SI{15}{\micro\second} of each cycle.
Table~\ref{Tab:FreqDetails} lists all settings relevant to the laser frequencies.

\begin{table}[htp]
\centering
\caption{Intermediate frequencies to determine the laser frequencies in the trap (see Fig.~\ref{Fig:LaserLock}).
During measurements the repetition rate of the comb $\nu_\text{comb,rep}$ was varied to change $\nu_1$.}
\label{Tab:FreqDetails}
\begin{ruledtabular}
\begin{tabular}{l r}
Frequency & Value \\
\hline
$\nu_\text{comb,rep}$ & \SI{250000233.5}{\hertz} \\
$\nu_\text{comb,offset}$ & \SI{-40000000.0}{\hertz} \\
\hline
For light at wavelength $\lambda_1$: \rule{0pt}{9pt} \\
\multicolumn{2}{l}{$\nu_1 = 2 \times (\nu_\text{comb,offset} + m_1 \times \nu_\text{comb,rep} + \nu_{\text{B}_1}) + \nu_\text{AOM-1}$} \\
$m_1$ (mode number) & 1214851 \\
$\nu_{\text{B}_1}$ & \SI{+29.01(1)}{\mega\hertz} \\
$\nu_\text{AOM-1}$ & \SI{+198.90}{\mega\hertz} \\
\hline
For light at wavelength $\lambda_2$: \rule{0pt}{9pt} \\
\multicolumn{2}{l}{$\nu_2 = \nu_\text{comb,offset} + m_2 \times \nu_\text{comb,rep} + \nu_{\text{B}_2} + \nu_{\text{B}'_2} + 2 \times \nu_\text{AOM-2}$} \\
$m_2$ (mode number) & 1845248 \\
$\nu_{\text{B}_2}$ & \SI{-27.33(1)}{\mega\hertz} \\
$\nu_{\text{B}'_2}$ & $-[\num{1116.8(1)}\text{--}\num{1274.2(1)}]$\,\si{\mega\hertz} \\
$\nu_\text{AOM-2}$ (double pass) & \SI{+348.00}{\mega\hertz} \\
\end{tabular}
\end{ruledtabular}
\end{table}


The Doppler-free spectrum of a single $\text{Ba}^\text{+}$ ion can be calculated
by solving the Liouville equation for the eight-level system of Fig.~\ref{Fig:BaLevels},
\begin{equation*}
\frac{d}{dt}\rho_{ij}=\frac{i}{\hbar}\sum_k (H_{ik}\rho_{kj}-\rho_{ik}H_{kj}) + 
\mathcal{R}_{ij}(\rho)\text{,}
\end{equation*}
where $H$ is the Hamiltonian describing the interaction with 
two laser fields and $\mathcal{R}$ is the damping matrix modeling relaxation and decoherence phenomena.
The magnetic field defines the $z$ direction
and both light fields are taken to propagate along the $-y$ direction (see Fig.~\ref{Fig:Trap}).
For the measurements presented herein, the light at wavelength $\lambda_1$ is linearly polarized, parallel to the magnetic field direction,
and the light at wavelength $\lambda_2$ is circularly polarized. The Hamiltonian that describes 
the coupling of the eight-level system (ordered according to Fig~\ref{Fig:BaLevels}) to these two light fields in the rotating wave approximation
is given by the $8\times8$ matrix
\begin{widetext}
\begin{equation*}
H = \hbar\left( \begin{array}{ccccccccc}
\Delta_1-\omega_B & 0 & -\frac{2}{\sqrt{3}}\Omega_1 & 0 & 0 & 0 & 0 & 0 \\
0 & \Delta_1+\omega_B & 0 & \frac{2}{\sqrt{3}}\Omega_1 & 0 & 0 & 0 & 0 \\
-\frac{2}{\sqrt{3}}\Omega_1 & 0 & -\frac{1}{3}\omega_B & 0 & \frac{i}{\sqrt{2}}\Omega_2 & \frac{2}{\sqrt{6}}\Omega_2 & -\frac{i}{\sqrt{6}}\Omega_2 & 0 \\
0 & \frac{2}{\sqrt{3}}\Omega_1 & 0 & \frac{1}{3}\omega_B & 0 & \frac{i}{\sqrt{6}}\Omega_2 & \frac{2}{\sqrt{6}}\Omega_2 & -\frac{i}{\sqrt{2}}\Omega_2 \\
0 & 0 & -\frac{i}{\sqrt{2}}\Omega_2 & 0 & \Delta_2-\frac{6}{5}\omega_B & 0 & 0 & 0 \\
0 & 0 & \frac{2}{\sqrt{6}}\Omega_2 & -\frac{i}{\sqrt{6}}\Omega_2 & 0 & \Delta_2-\frac{2}{5}\omega_B & 0 & 0 \\
0 & 0 & \frac{i}{\sqrt{6}}\Omega_2 & \frac{2}{\sqrt{6}}\Omega_2 & 0 & 0 & \Delta_2+\frac{2}{5}\omega_B & 0 \\
0 & 0 & 0 & \frac{i}{\sqrt{2}}\Omega_2 & 0 & 0 & 0 & \Delta_2+\frac{6}{5}\omega_B \\
\end{array} \right) \text{,}
\end{equation*}
\end{widetext}
where $\omega_B=\mu_B|\bvec{B}|/\hbar$ is the Larmor frequency, $\Delta_1$ and 
$\Delta_2$ are the detunings of the two laser frequencies with respect to
the $\text{Ba}^\text{+}$ transitions, 
and $\Omega_1$ and $\Omega_2$ are the corresponding Rabi frequencies,
depending on the amplitudes of the laser fields $\mathcal{E}_1$ and $\mathcal{E}_2$ as
\begin{eqnarray*}
\Omega_1 &=& \frac{1}{2\hbar} \mathcal{E}_1 \langle \slevel{} \Vert er \Vert \plevel{} \rangle \\
\Omega_2 &=& \frac{1}{2\hbar} \mathcal{E}_2 \langle \plevel{} \Vert er \Vert \dlevel{} \rangle \text{.}
\end{eqnarray*}

The relaxation matrix $\mathcal{R}(\rho)$ includes the spontaneous decay of the 
\plevel{} level and the decoherence effect due to finite laser linewidths.
Here $\Gamma_1 = \SI{14.7}{\mega\hertz}$ and $\Gamma_2 = \SI{5.4}{\mega\hertz}$ are the partial decay rates of the \plevel{} -- 
\slevel{} and \plevel{} -- \dlevel{} transitions~\cite{DeMunshi2014,Kuske1978}, such that the total decay rate is $\Gamma = \Gamma_1 + \Gamma_2$
with associated decoherence rate $\gamma = \Gamma/2$. The linewidths of the two lasers are both taken to be equal to 
$\gamma_l$. Using $\gamma' = \gamma + \gamma_l$, the total relaxation matrix is given by
\begin{widetext}
\begin{equation*}
\mathcal{R} = \left( \begin{array}{cccccccc}
\Gamma_{\!1}(\frac{1}{3}\rho_{33}+\frac{2}{3}\rho_{44}) & 
-\Gamma_{\!1}\frac{1}{3}\rho_{34} & -\gamma'\rho_{13} & -\gamma'\rho_{14} & 
-\gamma_l\rho_{15} & -\gamma_l\rho_{16} & -\gamma_l\rho_{17} & 
-\gamma_l\rho_{18} \\
-\Gamma_{\!1}\frac{1}{3}\rho_{43} & 
\Gamma_{\!1}(\frac{2}{3}\rho_{33}+\frac{1}{3}\rho_{44}) & -\gamma'\rho_{23} & 
-\gamma'\rho_{24} & -\gamma_l\rho_{25} & -\gamma_l\rho_{26} & -\gamma_l\rho_{27} 
& -\gamma_l\rho_{28} \\
-\gamma'\rho_{31} & -\gamma'\rho_{32} & -\Gamma\rho_{33} & -\Gamma\rho_{34} & 
-\gamma'\rho_{35} & -\gamma'\rho_{36} & -\gamma'\rho_{37} & -\gamma'\rho_{38} \\
-\gamma'\rho_{41} & -\gamma'\rho_{42} & -\Gamma\rho_{43} & -\Gamma\rho_{44} & 
-\gamma'\rho_{45} & -\gamma'\rho_{46} & -\gamma'\rho_{47} & -\gamma'\rho_{48} \\
-\gamma_l\rho_{51} & -\gamma_l\rho_{52} & -\gamma'\rho_{53} & -\gamma'\rho_{54} 
& \Gamma_{\!2}\frac{1}{2}\rho_{33} & \Gamma_{\!2}\frac{1}{2\sqrt{3}}\rho_{34} & 
0 & 0 \\
-\gamma_l\rho_{61} & -\gamma_l\rho_{62} & -\gamma'\rho_{63} & -\gamma'\rho_{64} 
& \Gamma_{\!2}\frac{1}{2\sqrt{3}}\rho_{43} & 
\Gamma_{\!2}(\frac{1}{3}\rho_{33}+\frac{1}{6}\rho_{44}) & 
\Gamma_{\!2}\frac{1}{3}\rho_{34} & 0 \\
-\gamma_l\rho_{71} & -\gamma_l\rho_{72} & -\gamma'\rho_{73} & -\gamma'\rho_{74} 
& 0 & \Gamma_{\!2}\frac{1}{3}\rho_{43} & 
\Gamma_{\!2}(\frac{1}{6}\rho_{33}+\frac{1}{3}\rho_{44}) & 
\Gamma_{\!2}\frac{1}{2\sqrt{3}}\rho_{34} \\
-\gamma_l\rho_{81} & -\gamma_l\rho_{82} & -\gamma'\rho_{83} & -\gamma'\rho_{84} 
& 0 & 0 & \Gamma_{\!2}\frac{1}{2\sqrt{3}}\rho_{43} & 
\Gamma_{\!2}\frac{1}{2}\rho_{44} \\
\end{array} \right) \text{.}
\end{equation*}
\end{widetext}

The spectrum of a $\text{Ba}^\text{+}$ ion is 
recorded by measuring the fluorescence signal while scanning frequency $\nu_2$ across the \dlevel{} -- \plevel{} 
resonance. To fit this data, the population of the \pterm{} level ($\rho_{33} + \rho_{44}$)
is obtained from the steady-state solution of the optical Bloch equations while varying $\Delta_2$.
The detuning parameters $\Delta_1$ and $\Delta_2$,
\begin{eqnarray*}
 \Delta_1 &=& \nu_1 - \nusd - \nudp \\
 \Delta_2 &=& \nu_2 - \nudp \text{,}
\end{eqnarray*}
relate the known laser frequencies $\nu_1$ and $\nu_2$ to the $\text{Ba}^\text{+}$ transition frequencies.
The chosen parametrization minimizes the correlation between fit parameters.
A numerical $\chi^2$ minimization is performed fitting
transition frequencies $\nudp$, $\nusd$,
Rabi frequencies $\Omega_1$ and $\Omega_2$,
and laser linewidth $\gamma_l$ to background-subtracted data. Ion dynamics are not taken into account.

\begin{figure}
\begin{center}
   \includegraphics[width=\hsize]{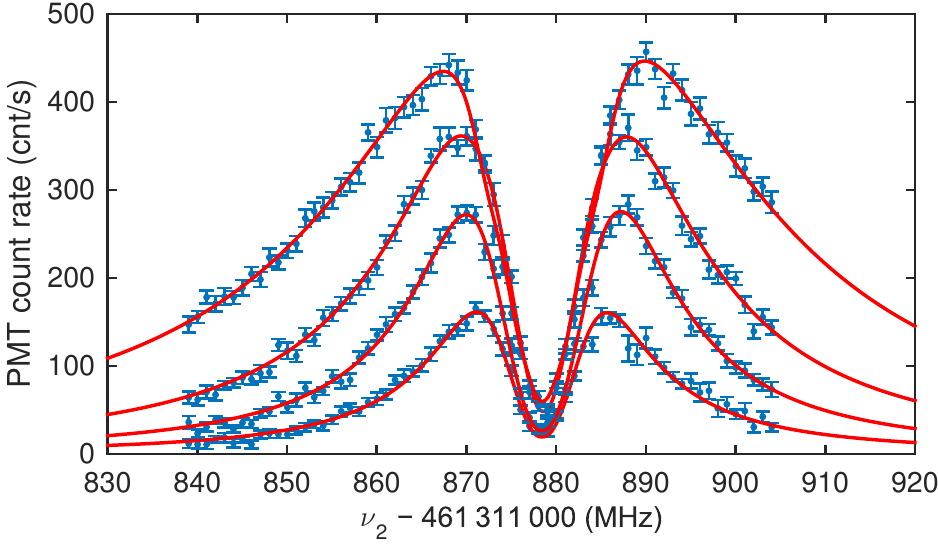}
  \caption{Spectra of the \dlevel{} -- \plevel{} transition in a single 
$^\text{138}\text{Ba}^\text{+}$ ion recorded for different light intensities $\Omega_2$, ranging from 0.3 to 4 times saturation intensity.
Frequency $\nu_1$ is kept constant with $\Delta_1 \approx 0$.
Solid lines correspond to the result of fitting the optical Bloch model to the data.
The width of the spectra show power broadening.}
  \label{Fig:PowerStack}
\end{center}
\end{figure}
\begin{figure}
\begin{center}
  \includegraphics[width=\hsize]{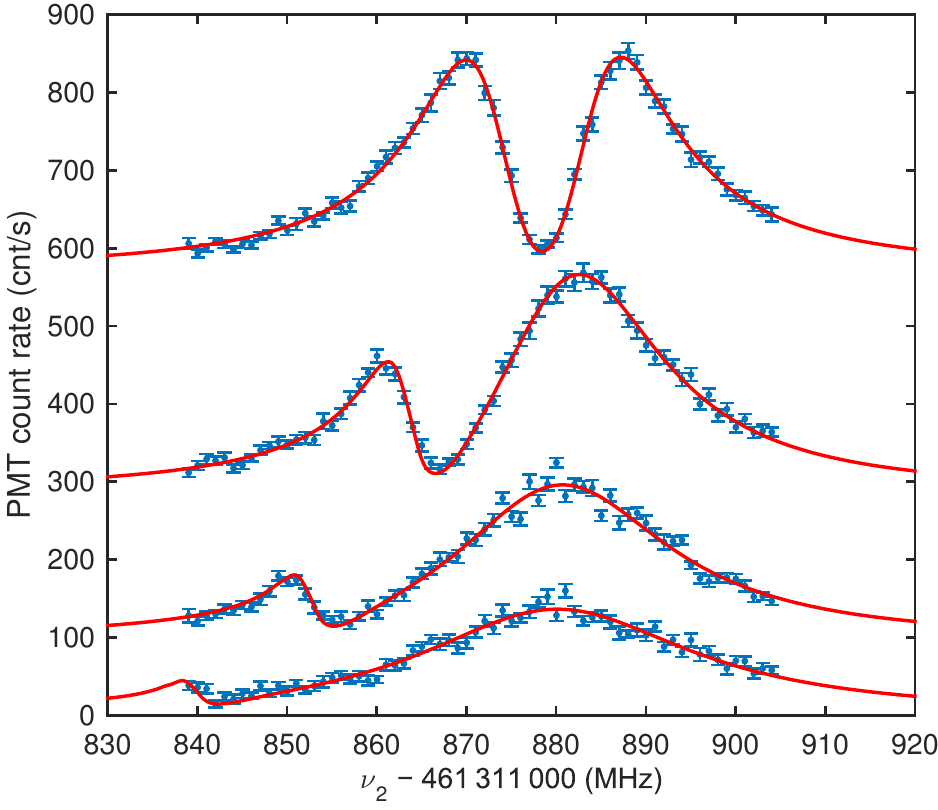}
  \caption{Spectra of the \dlevel{} -- \plevel{} transition in a single 
$^\text{138}\text{Ba}^\text{+}$ ion recorded for different detunings $\Delta_1$.
Note that the baselines are shifted to show the spectra.
Detuning $\Delta_1$ is varied in steps of \SI{12}{\mega\hertz}; intensities of the light fields are kept constant. 
Solid lines correspond to the result of fitting the optical Bloch model to the data.}
  \label{Fig:DetuningStack}
\end{center}
\end{figure}

Fig.~\ref{Fig:PowerStack} shows a set of four spectra recorded with different intensities of the light at wavelength $\lambda_2$,
corresponding to different Rabi frequencies $\Omega_2$.
The intensity ranged from $0.3$ to $4$ times the saturation intensity of the \dlevel{} -- \plevel{} transition.
Frequency $\nu_1$ was kept constant during these measurements.
The most prominent features in the spectra are the wide one-photon peak of the \dlevel{} -- \plevel{} transition centered at $\Delta_2 = 0$
and a dip in fluorescence (electromagnetically induced transparency) caused by the two-photon process at $\Delta_1 = \Delta_2$,
which is at the same frequency in this case since $\Delta_1 \approx 0$.
The polarizations of the two laser fields determine which coherences cause a fluorescence dip.
For the measurements presented here the dominant contributions are from coherences $\vert1\rangle\langle5\vert$ and $\vert2\rangle\langle8\vert$ (refer to Fig.~\ref{Fig:BaLevels});
the magnetic field magnitude is too small to resolve the two individual components here.
Fig.~\ref{Fig:DetuningStack} shows spectra recorded with constant laser intensities and different detunings $\Delta_1$.

\begin{figure}
\begin{center}
  \includegraphics[width=\hsize]{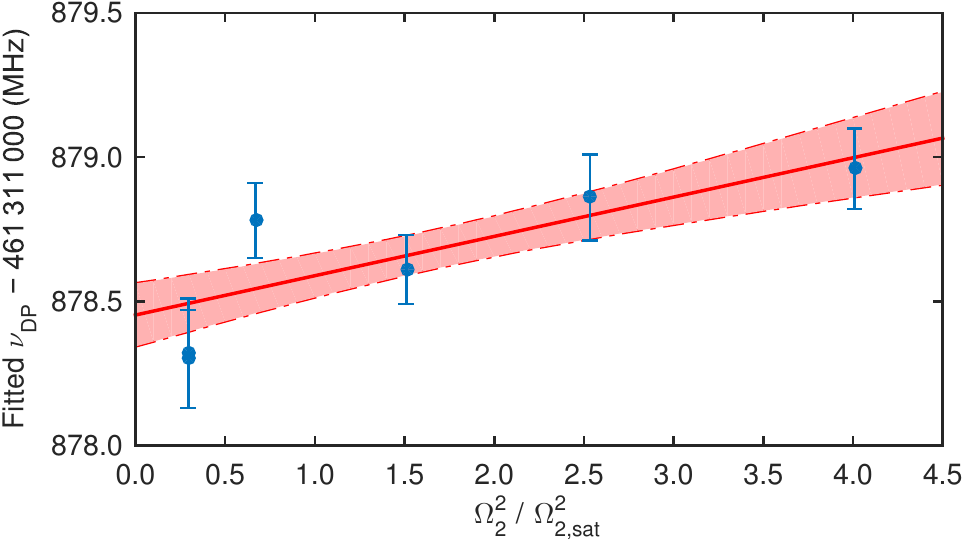}
  \caption{Extracted transition frequency $\nudp$ as a function of laser intensity $\Omega_2$ (given in terms of saturation intensity) for $\Delta_1 \approx 0$.
  The data shown include the spectra of Fig.~\ref{Fig:PowerStack}. An intensity-dependent shift can be seen,
  which is attributed to a small mismatch between model and experiment in a transition amplitude component that causes a light shift.
  The solid line corresponds to a linear extrapolation to zero laser light intensity, with $1\sigma$ confidence bounds indicated.}
  \label{Fig:ResidualShift}
\end{center}
\end{figure}

A line shape calculated by numerically solving the eight-level optical Bloch 
equations is fitted to each individual spectrum.
The scaling of the Rabi frequencies $\Omega_1$ and $\Omega_2$ with recorded laser powers,
and the laser linewidth $\gamma_l$ are taken from a global fit to all data.
Each of the spectra yields a value for the $\text{Ba}^\text{+}$ transition frequencies.
The fit results are found to be consistent, except for transition frequency $\nudp$.
Here a small dependence on the laser intensity corresponding to $\Omega_2$ is found as shown in Fig.~\ref{Fig:ResidualShift}.
Intensity-dependent light shift effects are included in the optical Bloch equations and will drop out in the fitting procedure.
However, a small mismatch in polarization or magnetic field orientation between model and experiment
can modify a transition amplitude component resulting in a mismatch in light shift.
The atomic transition frequency is determined by extrapolating the fitted value to zero laser intensity (see Fig.~\ref{Fig:ResidualShift}).
A second set of spectra recorded at detuning $\Delta_1 = \SI{-1.2}{\mega\hertz}$ yielded the same result.
The weighted average of the \dlevel{} -- \plevel{} and \slevel{} -- \dlevel{} transition frequencies
as well as their sum, the \slevel{} -- \plevel{} transition frequency, are given in Table~\ref{Tab:FinalFreqs}.
The results presented here are limited by statistics and the stability of the laser system.

\begin{table}[htp]
\centering
\caption{Transition frequencies of the \dlevel{} -- \plevel{} and \slevel{} -- \dlevel{} transitions in $^\text{138}\text{Ba}^\text{+}$.
Their sum yields the frequency of the \slevel{} -- \plevel{} transition.}
\label{Tab:FinalFreqs}
\begin{ruledtabular}
\begin{tabular}{lrr}
Transition & Frequency (MHz) & Relative uncertainty \\
\hline
$\nudp$ \rule[-5pt]{0pt}{14pt} & \num{461311878.5(1)} & $2 \times 10^{-10}$ \\
$\nusd$ \rule[-5pt]{0pt}{14pt} & \num{146114384.0(1)} & $6 \times 10^{-10}$ \\
\hline
$\nusp$ \rule[-5pt]{0pt}{14pt} & \num{607426262.5(2)} & $3 \times 10^{-10}$ \\
\end{tabular}
\end{ruledtabular}
\end{table}

\begin{figure}
\begin{center}
  \includegraphics[width=\hsize]{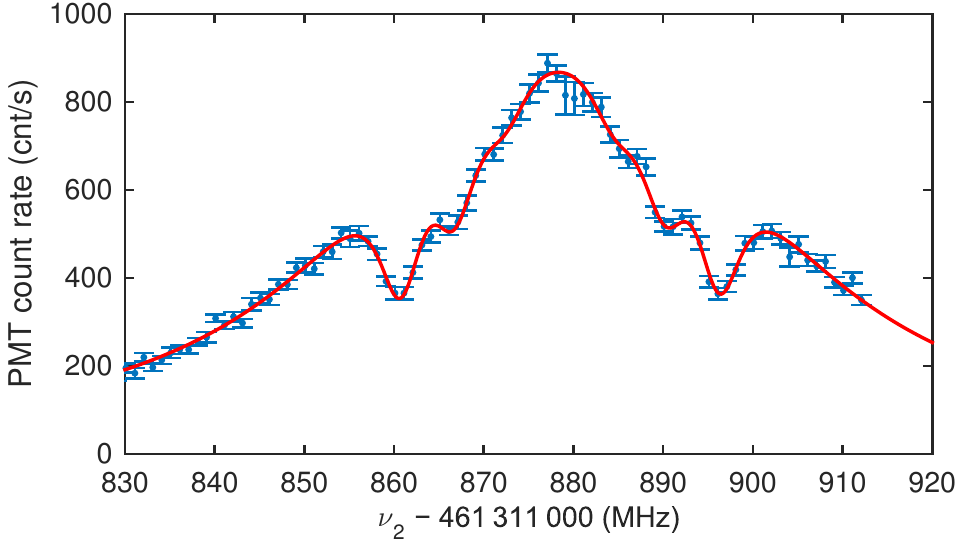}
  \caption{Spectrum of the \dlevel{} -- \plevel{} transition in a single $^\text{138}\text{Ba}^\text{+}$ ion.
  Both light fields are linearly polarized under an angle of \SI{75}{\degree} to the magnetic field direction,
  with about 15\% admixture of circularly polarized light.
  The solid line corresponds to the fit of the adjusted optical Bloch model, taking into account the polarization of the laser light.
  Detuning $\Delta_1 \approx 0$ results in the symmetric line shape. Even for a positive detuning $\Delta_1$ the ion can still be cooled and localized.}
  \label{Fig:TwoDips}
\end{center}
\end{figure}

Additional spectra have been recorded for different magnetic field settings, laser intensities,
and laser polarizations to study systematic effects. 
An example of such a spectrum is shown in Fig.~\ref{Fig:TwoDips} where both the light at wavelength $\lambda_1$ and $\lambda_2$ is 
about 85\% linearly polarized under an angle of \SI{75}{\degree} to a magnetic field of \SI{600}{\micro\tesla}.
In this more complex polarization state, multiple dips in the fluorescence appear.
The outermost features are due to coherences $\vert1\rangle\langle8\vert$ and $\vert2\rangle\langle5\vert$.
The frequency difference between these outer dips can be employed as a calibration of the magnetic field strength ($\omega_B$).

The transition frequencies of the \slevel{} -- \plevel{}, \dlevel{} -- \plevel{}, and \slevel{} -- \dlevel{} transitions
in $^\text{138}\text{Ba}^\text{+}$ have been determined to sub-MHz precision using a single trapped ion (see Table~\ref{Tab:FinalFreqs}).
An eight-level optical Bloch model describes well the line shapes under the conditions in the experiment.
Previous measurements had obtained these transition frequencies between low-lying levels to order \SI{100}{\mega\hertz} accuracy
using Fourier transform spectroscopy in large ion samples~\cite{Karlsson1999,Curry2004}.
This work improves upon the earlier measurements by more than two orders of magnitude.

\begin{acknowledgments}
The authors would like to acknowledge technical support by O.~B\"oll and L.~Huisman. This 
work is supported by the Foundation for Fundamental Research on Matter (FOM), 
which is part of the Netherlands Organization for Scientific Research (NWO), 
under programme 114 \emph{TRI\textmu{}P} and programme 125 \emph{Broken Mirrors 
\& Drifting Constants}.
\end{acknowledgments}

\bibliography{SpecPaper}

\begin{thebibliography}{20}%
\makeatletter
\providecommand \@ifxundefined [1]{%
 \@ifx{#1\undefined}
}%
\providecommand \@ifnum [1]{%
 \ifnum #1\expandafter \@firstoftwo
 \else \expandafter \@secondoftwo
 \fi
}%
\providecommand \@ifx [1]{%
 \ifx #1\expandafter \@firstoftwo
 \else \expandafter \@secondoftwo
 \fi
}%
\providecommand \natexlab [1]{#1}%
\providecommand \enquote  [1]{``#1''}%
\providecommand \bibnamefont  [1]{#1}%
\providecommand \bibfnamefont [1]{#1}%
\providecommand \citenamefont [1]{#1}%
\providecommand \href@noop [0]{\@secondoftwo}%
\providecommand \href [0]{\begingroup \@sanitize@url \@href}%
\providecommand \@href[1]{\@@startlink{#1}\@@href}%
\providecommand \@@href[1]{\endgroup#1\@@endlink}%
\providecommand \@sanitize@url [0]{\catcode `\\12\catcode `\$12\catcode
  `\&12\catcode `\#12\catcode `\^12\catcode `\_12\catcode `\%12\relax}%
\providecommand \@@startlink[1]{}%
\providecommand \@@endlink[0]{}%
\providecommand \url  [0]{\begingroup\@sanitize@url \@url }%
\providecommand \@url [1]{\endgroup\@href {#1}{\urlprefix }}%
\providecommand \urlprefix  [0]{URL }%
\providecommand \Eprint [0]{\href }%
\providecommand \doibase [0]{http://dx.doi.org/}%
\providecommand \selectlanguage [0]{\@gobble}%
\providecommand \bibinfo  [0]{\@secondoftwo}%
\providecommand \bibfield  [0]{\@secondoftwo}%
\providecommand \translation [1]{[#1]}%
\providecommand \BibitemOpen [0]{}%
\providecommand \bibitemStop [0]{}%
\providecommand \bibitemNoStop [0]{.\EOS\space}%
\providecommand \EOS [0]{\spacefactor3000\relax}%
\providecommand \BibitemShut  [1]{\csname bibitem#1\endcsname}%
\let\auto@bib@innerbib\@empty
\bibitem [{\citenamefont {Ludlow}\ \emph {et~al.}(2015)\citenamefont {Ludlow},
  \citenamefont {Boyd}, \citenamefont {Ye}, \citenamefont {Peik},\ and\
  \citenamefont {Schmidt}}]{Ludlow2015}%
  \BibitemOpen
  \bibfield  {author} {\bibinfo {author} {\bibfnamefont {A.~D.}\ \bibnamefont
  {Ludlow}}, \bibinfo {author} {\bibfnamefont {M.~M.}\ \bibnamefont {Boyd}},
  \bibinfo {author} {\bibfnamefont {J.}~\bibnamefont {Ye}}, \bibinfo {author}
  {\bibfnamefont {E.}~\bibnamefont {Peik}}, \ and\ \bibinfo {author}
  {\bibfnamefont {P.~O.}\ \bibnamefont {Schmidt}},\ }\href@noop {} {\
  (\bibinfo {year} {2015})},\ \Eprint {http://arxiv.org/abs/1407.3493}
  {arXiv:1407.3493 [physics.atom-ph]} \BibitemShut {NoStop}%
\bibitem [{\citenamefont {Hall}(2006)}]{Hall2006}%
  \BibitemOpen
  \bibfield  {author} {\bibinfo {author} {\bibfnamefont {J.~L.}\ \bibnamefont
  {Hall}},\ }\href {\doibase 10.1103/RevModPhys.78.1279} {\bibfield  {journal}
  {\bibinfo  {journal} {Rev. Mod. Phys.}\ }\textbf {\bibinfo {volume} {78}},\
  \bibinfo {pages} {1279} (\bibinfo {year} {2006})}\BibitemShut {NoStop}%
\bibitem [{\citenamefont {H\"ansch}(2006)}]{Haensch2006}%
  \BibitemOpen
  \bibfield  {author} {\bibinfo {author} {\bibfnamefont {T.~W.}\ \bibnamefont
  {H\"ansch}},\ }\href {\doibase 10.1103/RevModPhys.78.1297} {\bibfield
  {journal} {\bibinfo  {journal} {Rev. Mod. Phys.}\ }\textbf {\bibinfo {volume}
  {78}},\ \bibinfo {pages} {1297} (\bibinfo {year} {2006})}\BibitemShut
  {NoStop}%
\bibitem [{\citenamefont {Fortson}(1993)}]{Fortson1993}%
  \BibitemOpen
  \bibfield  {author} {\bibinfo {author} {\bibfnamefont {N.}~\bibnamefont
  {Fortson}},\ }\href {\doibase 10.1103/PhysRevLett.70.2383} {\bibfield
  {journal} {\bibinfo  {journal} {Phys. Rev. Lett.}\ }\textbf {\bibinfo
  {volume} {70}},\ \bibinfo {pages} {2383} (\bibinfo {year}
  {1993})}\BibitemShut {NoStop}%
\bibitem [{\citenamefont {Koerber}\ \emph {et~al.}(2003)\citenamefont
  {Koerber}, \citenamefont {Schacht}, \citenamefont {Nagourney},\ and\
  \citenamefont {Fortson}}]{Koerber2003}%
  \BibitemOpen
  \bibfield  {author} {\bibinfo {author} {\bibfnamefont {T.~W.}\ \bibnamefont
  {Koerber}}, \bibinfo {author} {\bibfnamefont {M.}~\bibnamefont {Schacht}},
  \bibinfo {author} {\bibfnamefont {W.}~\bibnamefont {Nagourney}}, \ and\
  \bibinfo {author} {\bibfnamefont {E.~N.}\ \bibnamefont {Fortson}},\ }\href
  {\doibase 10.1088/0953-4075/36/3/320} {\bibfield  {journal} {\bibinfo
  {journal} {J. Phys. B}\ }\textbf {\bibinfo {volume} {36}},\ \bibinfo {pages}
  {637} (\bibinfo {year} {2003})}\BibitemShut {NoStop}%
\bibitem [{\citenamefont {Wansbeek}\ \emph {et~al.}(2008)\citenamefont
  {Wansbeek}, \citenamefont {Sahoo}, \citenamefont {Timmermans}, \citenamefont
  {Jungmann}, \citenamefont {Das},\ and\ \citenamefont
  {Mukherjee}}]{Wansbeek2008}%
  \BibitemOpen
  \bibfield  {author} {\bibinfo {author} {\bibfnamefont {L.~W.}\ \bibnamefont
  {Wansbeek}}, \bibinfo {author} {\bibfnamefont {B.~K.}\ \bibnamefont {Sahoo}},
  \bibinfo {author} {\bibfnamefont {R.~G.~E.}\ \bibnamefont {Timmermans}},
  \bibinfo {author} {\bibfnamefont {K.}~\bibnamefont {Jungmann}}, \bibinfo
  {author} {\bibfnamefont {B.~P.}\ \bibnamefont {Das}}, \ and\ \bibinfo
  {author} {\bibfnamefont {D.}~\bibnamefont {Mukherjee}},\ }\href {\doibase
  10.1103/PhysRevA.78.050501} {\bibfield  {journal} {\bibinfo  {journal} {Phys.
  Rev. A}\ }\textbf {\bibinfo {volume} {78}},\ \bibinfo {pages} {050501(R)}
  (\bibinfo {year} {2008})}\BibitemShut {NoStop}%
\bibitem [{\citenamefont {{Nu\~nez Portela}}\ \emph {et~al.}(2014)\citenamefont
  {{Nu\~nez Portela}}, \citenamefont {Dijck}, \citenamefont {Mohanty},
  \citenamefont {Bekker}, \citenamefont {van~den Berg}, \citenamefont {Giri},
  \citenamefont {Hoekstra}, \citenamefont {Onderwater}, \citenamefont
  {Schlesser}, \citenamefont {Timmermans}, \citenamefont {Versolato},
  \citenamefont {Willmann}, \citenamefont {Wilschut},\ and\ \citenamefont
  {Jungmann}}]{NunezPortela2014}%
  \BibitemOpen
  \bibfield  {author} {\bibinfo {author} {\bibfnamefont {M.}~\bibnamefont
  {{Nu\~nez Portela}}}, \bibinfo {author} {\bibfnamefont {E.~A.}\ \bibnamefont
  {Dijck}}, \bibinfo {author} {\bibfnamefont {A.}~\bibnamefont {Mohanty}},
  \bibinfo {author} {\bibfnamefont {H.}~\bibnamefont {Bekker}}, \bibinfo
  {author} {\bibfnamefont {J.~E.}\ \bibnamefont {van~den Berg}}, \bibinfo
  {author} {\bibfnamefont {G.~S.}\ \bibnamefont {Giri}}, \bibinfo {author}
  {\bibfnamefont {S.}~\bibnamefont {Hoekstra}}, \bibinfo {author}
  {\bibfnamefont {C.~J.~G.}\ \bibnamefont {Onderwater}}, \bibinfo {author}
  {\bibfnamefont {S.}~\bibnamefont {Schlesser}}, \bibinfo {author}
  {\bibfnamefont {R.~G.~E.}\ \bibnamefont {Timmermans}}, \bibinfo {author}
  {\bibfnamefont {O.~O.}\ \bibnamefont {Versolato}}, \bibinfo {author}
  {\bibfnamefont {L.}~\bibnamefont {Willmann}}, \bibinfo {author}
  {\bibfnamefont {H.~W.}\ \bibnamefont {Wilschut}}, \ and\ \bibinfo {author}
  {\bibfnamefont {K.}~\bibnamefont {Jungmann}},\ }\href {\doibase
  10.1007/s00340-013-5603-2} {\bibfield  {journal} {\bibinfo  {journal} {Appl.
  Phys. B}\ }\textbf {\bibinfo {volume} {114}},\ \bibinfo {pages} {173}
  (\bibinfo {year} {2014})}\BibitemShut {NoStop}%
\bibitem [{\citenamefont {Kumar}\ \emph {et~al.}(2013)\citenamefont {Kumar},
  \citenamefont {Mantry}, \citenamefont {Marciano},\ and\ \citenamefont
  {Souder}}]{Kumar2013}%
  \BibitemOpen
  \bibfield  {author} {\bibinfo {author} {\bibfnamefont {K.~S.}\ \bibnamefont
  {Kumar}}, \bibinfo {author} {\bibfnamefont {S.}~\bibnamefont {Mantry}},
  \bibinfo {author} {\bibfnamefont {W.~J.}\ \bibnamefont {Marciano}}, \ and\
  \bibinfo {author} {\bibfnamefont {P.~A.}\ \bibnamefont {Souder}},\ }\href
  {\doibase 10.1146/annurev-nucl-102212-170556} {\bibfield  {journal} {\bibinfo
   {journal} {Annu. Rev. Nucl. Part. Sci.}\ }\textbf {\bibinfo {volume} {63}},\
  \bibinfo {pages} {237} (\bibinfo {year} {2013})}\BibitemShut {NoStop}%
\bibitem [{\citenamefont {{De Munshi}}\ \emph {et~al.}(2014)\citenamefont {{De
  Munshi}}, \citenamefont {Dutta}, \citenamefont {Rebhi},\ and\ \citenamefont
  {Mukherjee}}]{DeMunshi2014}%
  \BibitemOpen
  \bibfield  {author} {\bibinfo {author} {\bibfnamefont {D.}~\bibnamefont {{De
  Munshi}}}, \bibinfo {author} {\bibfnamefont {T.}~\bibnamefont {Dutta}},
  \bibinfo {author} {\bibfnamefont {R.}~\bibnamefont {Rebhi}}, \ and\ \bibinfo
  {author} {\bibfnamefont {M.}~\bibnamefont {Mukherjee}},\ }\href@noop {} {\
  (\bibinfo {year} {2014})},\ \Eprint {http://arxiv.org/abs/1411.5041}
  {arXiv:1411.5041 [physics.atom-ph]} \BibitemShut {NoStop}%
\bibitem [{\citenamefont {Stalgies}\ \emph {et~al.}(1998)\citenamefont
  {Stalgies}, \citenamefont {Siemers}, \citenamefont {Appasamy},\ and\
  \citenamefont {Toschek}}]{Stalgies1998}%
  \BibitemOpen
  \bibfield  {author} {\bibinfo {author} {\bibfnamefont {Y.}~\bibnamefont
  {Stalgies}}, \bibinfo {author} {\bibfnamefont {I.}~\bibnamefont {Siemers}},
  \bibinfo {author} {\bibfnamefont {B.}~\bibnamefont {Appasamy}}, \ and\
  \bibinfo {author} {\bibfnamefont {P.~E.}\ \bibnamefont {Toschek}},\ }\href
  {\doibase 10.1364/JOSAB.15.002505} {\bibfield  {journal} {\bibinfo  {journal}
  {J. Opt. Soc. Am. B}\ }\textbf {\bibinfo {volume} {15}},\ \bibinfo {pages}
  {2505} (\bibinfo {year} {1998})}\BibitemShut {NoStop}%
\bibitem [{\citenamefont {Oberst}(1999)}]{Oberst1999}%
  \BibitemOpen
  \bibfield  {author} {\bibinfo {author} {\bibfnamefont {H.}~\bibnamefont
  {Oberst}},\ }\emph {\bibinfo {title} {Resonance Fluorescence of Single Barium
  Ions}},\ \href@noop {} {Master's thesis},\ \bibinfo  {school} {Universit\"at
  Innsbruck}, \bibinfo {address} {Innsbruck} (\bibinfo {year}
  {1999})\BibitemShut {NoStop}%
\bibitem [{\citenamefont {Zanon-Willette}\ \emph {et~al.}(2011)\citenamefont
  {Zanon-Willette}, \citenamefont {de~Clercq},\ and\ \citenamefont
  {Arimondo}}]{Zanon-Willette2011}%
  \BibitemOpen
  \bibfield  {author} {\bibinfo {author} {\bibfnamefont {T.}~\bibnamefont
  {Zanon-Willette}}, \bibinfo {author} {\bibfnamefont {E.}~\bibnamefont
  {de~Clercq}}, \ and\ \bibinfo {author} {\bibfnamefont {E.}~\bibnamefont
  {Arimondo}},\ }\href {\doibase 10.1103/PhysRevA.84.062502} {\bibfield
  {journal} {\bibinfo  {journal} {Phys. Rev. A}\ }\textbf {\bibinfo {volume}
  {84}},\ \bibinfo {pages} {062502} (\bibinfo {year} {2011})}\BibitemShut
  {NoStop}%
\bibitem [{\citenamefont {Siemers}\ \emph {et~al.}(1992)\citenamefont
  {Siemers}, \citenamefont {Schubert}, \citenamefont {Blatt}, \citenamefont
  {Neuhauser},\ and\ \citenamefont {Toschek}}]{Siemers1992}%
  \BibitemOpen
  \bibfield  {author} {\bibinfo {author} {\bibfnamefont {I.}~\bibnamefont
  {Siemers}}, \bibinfo {author} {\bibfnamefont {M.}~\bibnamefont {Schubert}},
  \bibinfo {author} {\bibfnamefont {R.}~\bibnamefont {Blatt}}, \bibinfo
  {author} {\bibfnamefont {W.}~\bibnamefont {Neuhauser}}, \ and\ \bibinfo
  {author} {\bibfnamefont {P.~E.}\ \bibnamefont {Toschek}},\ }\href {\doibase
  10.1209/0295-5075/18/2/009} {\bibfield  {journal} {\bibinfo  {journal}
  {Europhys. Lett.}\ }\textbf {\bibinfo {volume} {18}},\ \bibinfo {pages} {139}
  (\bibinfo {year} {1992})}\BibitemShut {NoStop}%
\bibitem [{\citenamefont {Pinkert}\ \emph {et~al.}(2015)\citenamefont
  {Pinkert}, \citenamefont {B\"oll}, \citenamefont {Willmann}, \citenamefont
  {Jansen}, \citenamefont {Dijck}, \citenamefont {Groeneveld}, \citenamefont
  {Smets}, \citenamefont {Bosveld}, \citenamefont {Ubachs}, \citenamefont
  {Jungmann}, \citenamefont {Eikema},\ and\ \citenamefont
  {Koelemeij}}]{Pinkert2015}%
  \BibitemOpen
  \bibfield  {author} {\bibinfo {author} {\bibfnamefont {T.~J.}\ \bibnamefont
  {Pinkert}}, \bibinfo {author} {\bibfnamefont {O.}~\bibnamefont {B\"oll}},
  \bibinfo {author} {\bibfnamefont {L.}~\bibnamefont {Willmann}}, \bibinfo
  {author} {\bibfnamefont {G.~S.~M.}\ \bibnamefont {Jansen}}, \bibinfo {author}
  {\bibfnamefont {E.~A.}\ \bibnamefont {Dijck}}, \bibinfo {author}
  {\bibfnamefont {B.~G. H.~M.}\ \bibnamefont {Groeneveld}}, \bibinfo {author}
  {\bibfnamefont {R.}~\bibnamefont {Smets}}, \bibinfo {author} {\bibfnamefont
  {F.~C.}\ \bibnamefont {Bosveld}}, \bibinfo {author} {\bibfnamefont
  {W.}~\bibnamefont {Ubachs}}, \bibinfo {author} {\bibfnamefont
  {K.}~\bibnamefont {Jungmann}}, \bibinfo {author} {\bibfnamefont {K.~S.~E.}\
  \bibnamefont {Eikema}}, \ and\ \bibinfo {author} {\bibfnamefont {J.~C.~J.}\
  \bibnamefont {Koelemeij}},\ }\href {\doibase 10.1364/AO.54.000728} {\bibfield
   {journal} {\bibinfo  {journal} {Appl. Opt.}\ }\textbf {\bibinfo {volume}
  {54}},\ \bibinfo {pages} {728} (\bibinfo {year} {2015})}\BibitemShut
  {NoStop}%
\bibitem [{\citenamefont {Dammalapati}\ \emph {et~al.}(2009)\citenamefont
  {Dammalapati}, \citenamefont {De}, \citenamefont {Jungmann},\ and\
  \citenamefont {Willmann}}]{Dammalapati2009}%
  \BibitemOpen
  \bibfield  {author} {\bibinfo {author} {\bibfnamefont {U.}~\bibnamefont
  {Dammalapati}}, \bibinfo {author} {\bibfnamefont {S.}~\bibnamefont {De}},
  \bibinfo {author} {\bibfnamefont {K.}~\bibnamefont {Jungmann}}, \ and\
  \bibinfo {author} {\bibfnamefont {L.}~\bibnamefont {Willmann}},\ }\href
  {\doibase 10.1140/epjd/e2009-00076-x} {\bibfield  {journal} {\bibinfo
  {journal} {Eur. Phys. J. D}\ }\textbf {\bibinfo {volume} {53}},\ \bibinfo
  {pages} {1} (\bibinfo {year} {2009})}\BibitemShut {NoStop}%
\bibitem [{\citenamefont {Gerstenkorn}\ and\ \citenamefont
  {Luc}(1980)}]{Gerstenkorn1980}%
  \BibitemOpen
  \bibfield  {author} {\bibinfo {author} {\bibfnamefont {S.}~\bibnamefont
  {Gerstenkorn}}\ and\ \bibinfo {author} {\bibfnamefont {P.}~\bibnamefont
  {Luc}},\ }\href@noop {} {\emph {\bibinfo {title} {Atlas du Spectre
  d'Absorption de la Mol\'ecule de d'Iode}}},\ Vol.\ \bibinfo {volume}
  {$14\,000\text{~cm}^{-1}$ -- $15\,600\text{~cm}^{-1}$}\ (\bibinfo
  {publisher} {Laboratoire Aim\'e Cotton, CNRS II, Orsay},\ \bibinfo {year}
  {1980})\BibitemShut {NoStop}%
\bibitem [{\citenamefont {Xu}\ \emph {et~al.}(2000)\citenamefont {Xu},
  \citenamefont {van Dierendonck}, \citenamefont {Hogervorst},\ and\
  \citenamefont {Ubachs}}]{Xu2000}%
  \BibitemOpen
  \bibfield  {author} {\bibinfo {author} {\bibfnamefont {S.~C.}\ \bibnamefont
  {Xu}}, \bibinfo {author} {\bibfnamefont {R.}~\bibnamefont {van Dierendonck}},
  \bibinfo {author} {\bibfnamefont {W.}~\bibnamefont {Hogervorst}}, \ and\
  \bibinfo {author} {\bibfnamefont {W.}~\bibnamefont {Ubachs}},\ }\href
  {\doibase 10.1006/jmsp.2000.8085} {\bibfield  {journal} {\bibinfo  {journal}
  {J. Mol. Spectrosc.}\ }\textbf {\bibinfo {volume} {201}},\ \bibinfo {pages}
  {256} (\bibinfo {year} {2000})}\BibitemShut {NoStop}%
\bibitem [{\citenamefont {Kuske}\ \emph {et~al.}(1978)\citenamefont {Kuske},
  \citenamefont {Kirchner}, \citenamefont {Wittmann}, \citenamefont {Andr\"a},\
  and\ \citenamefont {Kaiser}}]{Kuske1978}%
  \BibitemOpen
  \bibfield  {author} {\bibinfo {author} {\bibfnamefont {P.}~\bibnamefont
  {Kuske}}, \bibinfo {author} {\bibfnamefont {N.}~\bibnamefont {Kirchner}},
  \bibinfo {author} {\bibfnamefont {W.}~\bibnamefont {Wittmann}}, \bibinfo
  {author} {\bibfnamefont {H.~J.}\ \bibnamefont {Andr\"a}}, \ and\ \bibinfo
  {author} {\bibfnamefont {D.}~\bibnamefont {Kaiser}},\ }\href {\doibase
  10.1016/0375-9601(78)90271-2} {\bibfield  {journal} {\bibinfo  {journal}
  {Phys. Lett. A}\ }\textbf {\bibinfo {volume} {64}},\ \bibinfo {pages} {377}
  (\bibinfo {year} {1978})}\BibitemShut {NoStop}%
\bibitem [{\citenamefont {Karlsson}\ and\ \citenamefont
  {Litz\'en}(1999)}]{Karlsson1999}%
  \BibitemOpen
  \bibfield  {author} {\bibinfo {author} {\bibfnamefont {H.}~\bibnamefont
  {Karlsson}}\ and\ \bibinfo {author} {\bibfnamefont {U.}~\bibnamefont
  {Litz\'en}},\ }\href {\doibase 10.1238/Physica.Regular.060a00321} {\bibfield
  {journal} {\bibinfo  {journal} {Phys. Scr.}\ }\textbf {\bibinfo {volume}
  {60}},\ \bibinfo {pages} {321} (\bibinfo {year} {1999})}\BibitemShut
  {NoStop}%
\bibitem [{\citenamefont {Curry}(2004)}]{Curry2004}%
  \BibitemOpen
  \bibfield  {author} {\bibinfo {author} {\bibfnamefont {J.~J.}\ \bibnamefont
  {Curry}},\ }\href {\doibase 10.1063/1.1643404} {\bibfield  {journal}
  {\bibinfo  {journal} {J. Phys. Chem. Ref. Data}\ }\textbf {\bibinfo {volume}
  {33}},\ \bibinfo {pages} {725} (\bibinfo {year} {2004})}\BibitemShut
  {NoStop}%
\end{thebibliography}%

\end{document}